\documentclass[11pt,letterpaper]{article}
\usepackage[top=0.85in,left=0.75in,right=0.75in,footskip=0.75in]{geometry}
 \usepackage{multirow}
\usepackage{amsmath,amssymb}

\usepackage{changepage}

\usepackage[utf8x]{inputenc}

\usepackage{textcomp,marvosym}

\usepackage{cite}

\usepackage{nameref,hyperref}

\usepackage{microtype}
\DisableLigatures[f]{encoding = *, family = * }

\usepackage[table]{xcolor}

\usepackage{array}

\newcolumntype{+}{!{\vrule width 2pt}}

\newlength\savedwidth

\usepackage[aboveskip=1pt,labelfont=bf,labelsep=period,justification=raggedright,singlelinecheck=off]{caption}

\makeatletter
\renewcommand{\@biblabel}[1]{\quad#1.}
\makeatother

\usepackage{lastpage,fancyhdr,graphicx}
\usepackage{epstopdf}
\fancyheadoffset[L]{0.1in}
\fancyfootoffset[L]{0.1in}
\begin{document}
\vspace*{0.2in}
\begin{flushleft}
{\Large
\textbf\newline{Deep Learning Models May Spuriously Classify Covid-19 from X-ray Images Based on Confounders} 
}
\newline
\\
Kaoutar Ben Ahmed\textsuperscript{1,3,*},
Gregory M. Goldgof\textsuperscript{2,3},
Rahul Paul\textsuperscript{1},
Dmitry B. Goldgof\textsuperscript{1,4},
Lawrence O. Hall\textsuperscript{1,4}
\\
\bigskip
\textbf{1} Department of Computer Science and Engineering, University of South Florida, Tampa, USA 
\\
\textbf{2} Department of Laboratory Medicine, University of California, San Francisco, USA.
\\
\textbf{3} These authors contributed equally.
\\
\textbf{4} These authors jointly supervised.

\bigskip

%
%




* kbenahmed@usf.edu

\end{flushleft}
\section*{Abstract}

Identifying who is infected with the Covid-19 virus is critical for controlling its spread. X-ray machines are widely available worldwide and can quickly provide images that can be used for diagnosis. A number of recent studies claim it may be possible to build highly accurate  models, using deep learning, to detect Covid-19 from chest X-ray images. 
This paper explores the robustness and generalization ability of convolutional neural network models in diagnosing Covid-19 disease from frontal-view (AP/PA), raw chest X-ray images that were lung field cropped. Some concerning observations are made about high performing models that have learned to rely on confounding features related to the data source, rather than the patient's lung pathology, when differentiating between Covid-19 positive and negative labels. Specifically, these models likely made diagnoses based on confounding  factors such as patient age or image processing artifacts, rather than medically relevant information.

%


\section*{Introduction}

At the end of the year 2019, we witnessed the start of the ongoing global pandemic caused by Coronavirus disease (Covid-19) first identified in December 2019 in Wuhan, China. As of December 2020, more than 75 million cases are confirmed with more than 1.67 million confirmed deaths worldwide \cite{worldometers}.

In the first few months of the pandemic, the testing ability was limited in the US and other countries. Testing for Covid-19 has been unable to keep up with the demand at times and some tests required significant time to produce results (days) \cite{weissleder2020Covid}.

Therefore, other timely approaches to diagnosis were worthy of investigation \cite{jama}.
Chest X-rays (CXR) can be used to give relatively immediate diagnostic information.  X-ray machines are available in almost all diagnostic medical settings,  image acquisition is fast and relatively low cost. 

\begin{figure}[!h]
\caption{{\bf Modified Stop sign could be classified in a dangerous way. }}
\centering
\includegraphics[width=0.5\textwidth]{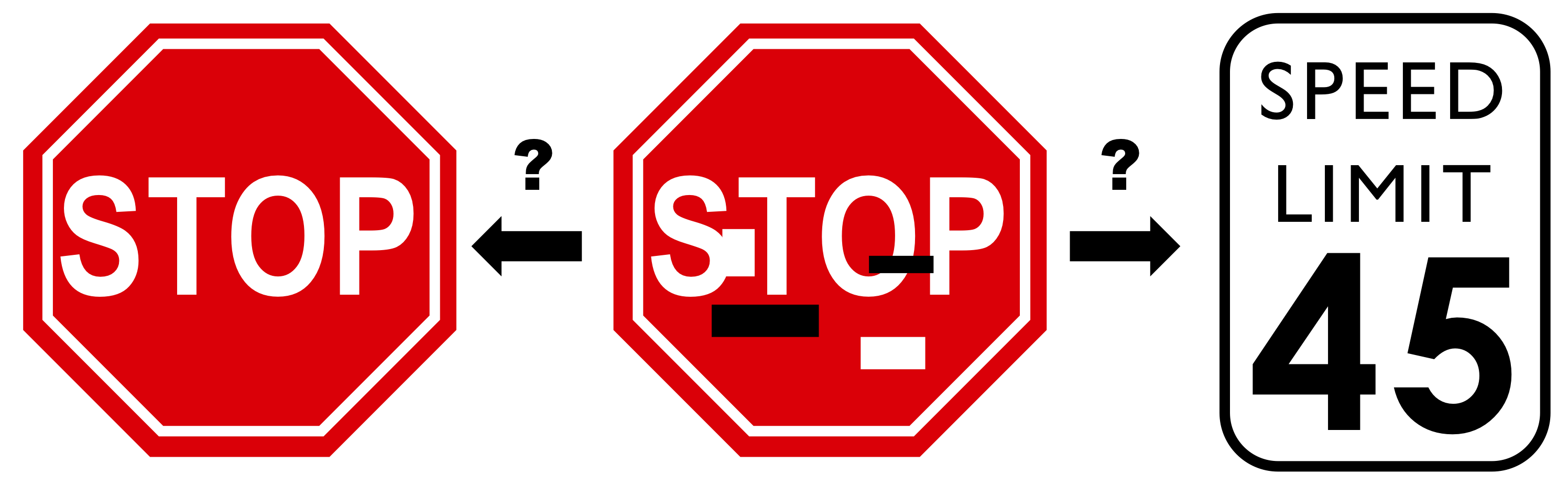}

\label{stop}
\end{figure}

Multiple studies were published claiming the possibility of diagnosing Covid-19 from chest X-rays using machine learning models with very high accuracy.
However, we believe that these models generalize very poorly and rely on learning shortcuts instead of true and relevant Covid-19 radiographic markers. These studies rely on deep learning approaches using convolutional neural networks (CNN) which automatically extract features.  A great concern with deep neural networks is whether the features they have learned for a particular problem are relevant. As an example, a study has shown that a CNN which learned to identify traffic signs will misclassify a stop sign as a 45 mile per hour speed limit sign, if just a couple of strips are placed on the sign without obscuring any text. This was demonstrated by the addition of a black or white sticker that did not obscure the 'STOP' word on the sign, a change that would have no effect on the human interpretation of the sign \cite{stopsign}.   Figure~\ref{stop} shows an example that we'd all interpret as a stop sign, but a CNN might misclassify. 

A recent survey \cite{chen2020survey} has discussed multiple recent papers applying Artificial Intelligence to chest imaging of Covid-19 (see Table \ref{survey}, where we have added new result rows and a column on testing data sources). In most of those papers authors used subsets of train/validation/test from the same source, others opted for a cross validation evaluation method, which also mixed train/val/test sources. In this paper, we show how the use of the same sources in train/test sets leads to the high accuracy that these models have achieved.
In addition, the majority of these papers used low quality images, some are cropped from PDF files of scientific publications. This approach increases the likelihood of introducing image processing artifacts, which further increases the risk of learning confounders rather than real pathologic features.

\begin{table}[!ht]
\centering 
\caption{
{\bf Papers for Automatic Covid-19 Prediction Based on CXR Images.}}
\begin{tabular}{|c|c|c|}
\hline \textbf{Paper} &  \textbf{Performance Result} &  \textbf{Testing Dataset}   \\ \hline
Ozturk et al. \cite{ozturk2020classification} & 0.99 AUC & Test on Training data\\ \hline
Abbas et al.  \cite{abbas2020classification} & 0.94 AUC & train/test split from the same  data source \\ \hline

Farooq et al.  \cite{farooq2020Covid} & 96.23\% Accuracy & train/test split from the same  data source \\ \hline
Lv et al.   \cite{lv2020cascade} & 85.62\% Accuracy & train/test split from the same  data source \\ \hline
Bassi and Attux   \cite{bassi2020deep} & 97.80\% Recall & train/test split from the same  data source \\ \hline
Rahimzadeh and Attar   \cite{rahimzadeh2020new} & 99.60\% Accuracy & train/test split from the same  data source \\ \hline
Chowdhury et al.    \cite{chowdhury2020can} & 98.30\% Accuracy & train/test split from the same  data source \\ \hline
Hemdan et al. \cite{hemdan2020Covidx} & 0.89 F1-score  & train/test split from the same  data source \\ \hline
Karim et al.  \cite{karim2020deepCovidexplainer} & 83.00\% Recall &  train/test split from the same  data source\\ \hline

Hall et al. \cite{hall2020finding} & 0.95 AUC  & 10-fold cross validation with all sources mixed \\ \hline
Apostolopoulos et al.  \cite{Apostolopoulos_2020} & 92.85\% Accuracy & 10-fold cross validation with all sources mixed\\ \hline
Apostolopoulos et al.  \cite{apostolopoulos2020extracting} & 99.18\% Accuracy & 10-fold cross validation with all sources mixed\\ \hline
Basu et al. \cite{basu2020deep} & 95.30\% Accuracy  & 5-fold cross validation with all sources mixed\\ \hline
Li et al.  \cite{li2020robust} & 97.01\% Accuracy  & 5-fold cross validation with all sources mixed\\ \hline

Yeh et al.    \cite{yeh2020cascaded} & 40\% Specificity & test on an unseen data source \\ \hline
This work & 0.96 AUC &  train/test split from the same  data source \\ \hline
This work & 0.63 AUC &  test on an unseen data source  \\ \hline

\end{tabular}
\label{survey}
\end{table}
Additionally, in some studies, \cite{panwar2020application,li2020robust,kassani2020automatic,oh2020deep,wang2020Covid} the pneumonia/normal class dataset was based on a pediatric dataset (age of patients 1-5 years of age). Whereas, the average age of the Covid-19 class was $>$40 years. By looking at the pneumonia image, it is evident that the sizes of the rib cages and thoracic structures of the pneumonia dataset are different from the Covid-19 cases, due to the age difference. These studies were likely using  age-related features to differentiate pneumonia/normal cases and Covid-19 cases, as a proxy for age rather than pathologic diagnosis, since convolutional neural networks have been shown to be able to learn the concept of size \cite{Cherezov-Size} (e.g. lung size).

In contrast, in Yeh et al. \cite{yeh2020cascaded}, authors have noticed that there is a generalization gap when testing on private datasets while the model was trained on open-source datasets. Furthermore, findings in the DeGrave et al. \cite{degrave2020ai} paper support our observations where authors investigated and showed, using saliency maps and generative adversarial  networks (GANs), that the model is actually learning shortcuts to differentiate between labels instead of Covid-19 pathology. In lay terms, this work demonstrated that the deep learning algorithms were looking at non-lung regions of the chest X-ray to classify the majority of images. 
The focus of this paper is to determine whether deep learning models can be considered reliable for diagnosing Covid-19 based on reasonable biomarkers, or are they only learning shortcuts (confounders) to differentiate between classes. To evaluate this question, we worked with 655 chest X-rays of patients diagnosed with Covid-19 and a set of 1,069 chest X-rays of patients diagnosed with other pneumonia that predates the emergence of Covid-19. 

 \subsection*{Contributions}
In our previous work \cite{hall2020finding}, we used Covid-19 images from three main sources \cite{cohen2020Covid}, \cite{radiopedia} and \cite{sirm}. Note that these sources were and still are largely used in the majority of research papers related to the prediction of Covid-19 from X-rays. We later identified a number of potential problems with these sources. Many of these images are extracted from PDF paper publications, are pre-processed with unknown methods, down-sampled, and are 3 channel (color). The exact source of the image is not always known and the stage of the disease is unknown.
Therefore, the main contributions of this paper are as follows:
(i) We found and use raw, high quality images from trustworthy sources.
(ii) The bias that might be introduced by the noise present around the corners of the images (dates, letters, arrows ...etc) was (mostly) removed by automatically segmenting the lung field and cropping the lung area based on a generated mask.
(iii) Finally, and most importantly, we discuss some troubling observations about the trained models and what features these models may be using to identify Covid-19 positivity/negativity.


\section*{Materials and Methods}

\subsection*{Datasets}

For the Covid-19 class, three sources were used in this work, BIMCV-Covid-19+ (Spain) \cite{de2020bimcv}, Covid-19-AR (USA) \cite{tcia_2020} and V2-COV19-NII (Germany) \cite{germanymeyer_2020}. For readability, we will label each dataset both by its name and also its country of origin, since the names of each dataset are similar and may confuse the reader. 

BIMCV Covid-19+ (Spain) is a large dataset from the Valencian Region Medical ImageBank (BIMCV) containing chest X-ray images CXR (CR, DX) and computed tomography (CT) imaging of Covid-19+ patients along with their radiological findings and locations, pathologies, radiological reports (in Spanish) and other data. The images provided are 16 bits in png format.

Covid-19-AR (USA) is a collection of radiographic (X-ray) and CT imaging studies of patients from The University of Arkansas for Medical Sciences Translational Research Institute who tested positive for Covid-19. Each patient is described by a limited set of clinical data  that includes demographics, comorbidities, selected lab data and key radiology findings. The provided images are in DICOM format.

V2-COV19-NII (Germany) is a repository containing image data collected by the Institute for Diagnostic and Interventional Radiology at the Hannover Medical School. It includes a dataset of Covid-19 cases with a focus on X-ray imaging. This includes images with extensive metadata, such as admission, ICU, laboratory, and anonymized patient data. The set contains raw, unprocessed, gray value image data as Nifti files.

Each patient in the datasets had different X-ray views (Lateral, AP or PA) and had multiple sessions of X-rays to assess the disease progress. Radiology reports and PCR test results were included in both BIMCV Covid-19+ and Covid-19-AR (USA) sources. We selected patients with AP and PA views. After translating and reading all the sessions reports coupled with PCR results, only one session per patient was chosen based on the disease stage. We picked the session with a positive PCR result and most severe stage.

\begin{figure}[!h]
\caption{{\bf Samples of the input X-rays. }
TOP: Covid-19 cases. BOTTOM: Pneumonia cases.}
\centering
\includegraphics[scale= 0.4]{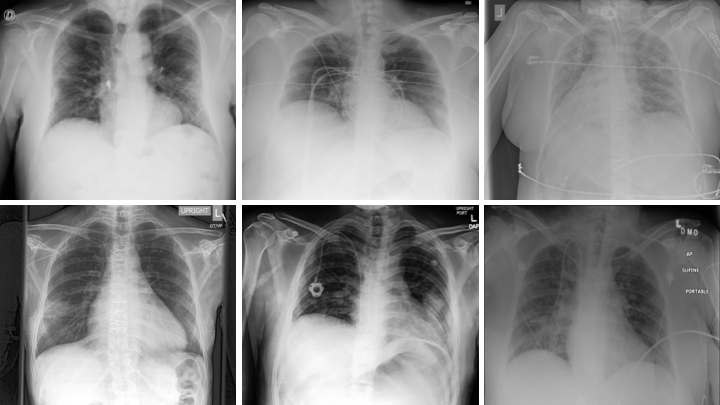}

\label{fig1}
\end{figure}

For the pneumonia class, We used 3 sources: (i) the National Institute of Health (NIH) dataset \cite{wang2017chestx}, (ii) Chexpert dataset \cite{irvin2019chexpert} and (iii) Padchest dataset \cite{bustos2020padchest}. The NIH and Chexpert dataset had pneumonia X-ray images with multiple labels (various lung disease conditions), but for simplicity, we chose the cases that had only one label (pneumonia). Only X-rays with a frontal view (AP or PA) were used in this work.
Three samples of Covid-19 and three pneumonia X-ray images are shown in Fig~\ref{fig1}.

\subsection*{Data Pre-processing}

\begin{figure}[!h]
\caption{{\bf Pipeline of Lung ROI cropping }}

\centering
\includegraphics[scale=0.6]{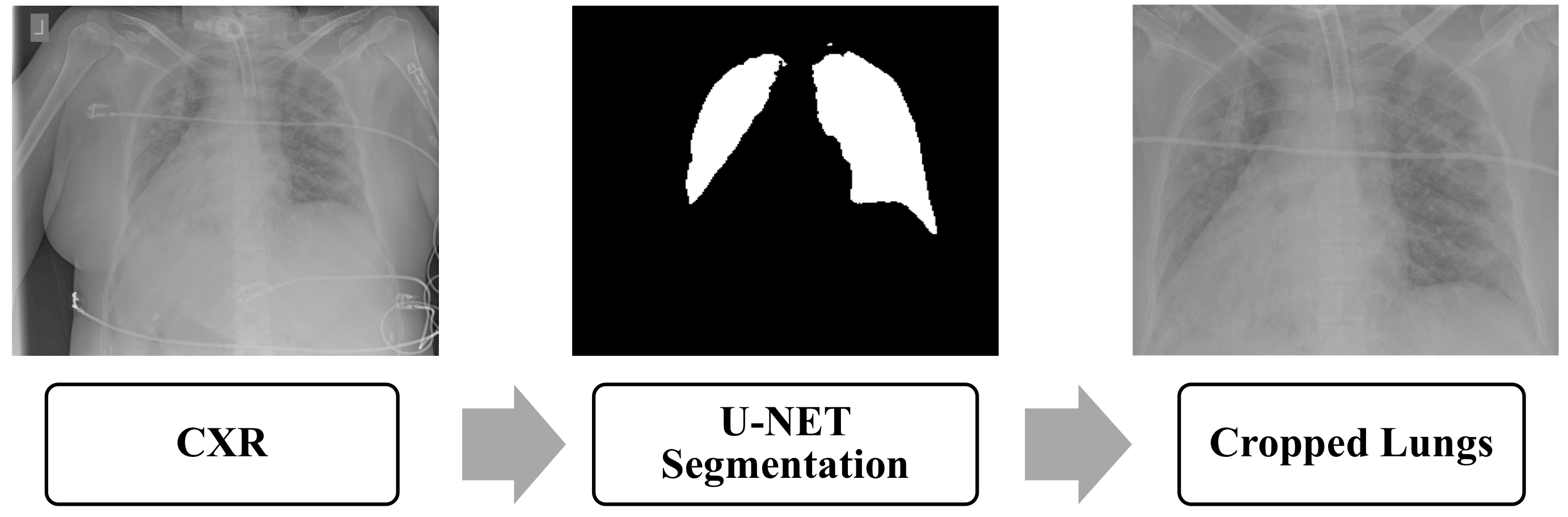}

\label{fig2}
\end{figure}

As stated in the previous section, the obtained images come in different formats.
Padchest \cite{bustos2019padchest} and BIMCV-Covid-19+ (Spain) \cite{de2020bimcv} datasets were processed by rescaling the dynamic range using the DICOM window width and center, when available. We do not know of any pre-processing steps applied to the other datasets.
As a first step we normalized all the images to 8 bits PNG format in the [0- 255] range.
The images were originally 1 grayscale channel, we duplicated them to 3 channels for use with  pre-trained deep neural networks.  The reason behind this is that Resnet50, the model that we utilized as a base model was pretrained on 8 bit color images. 
For lung ROI segmentation we used a UNET model pre-trained \cite{sivaramakrishnan2004iteratively} on a collection of CXRs with lung masks. The model generates 256x256 masks. We adapted their open source code \cite{sivaramakrishnan2004iteratively} to crop bounding boxes containing the lung area based on the generated masks. We resized the masks to the original input image size. We then added the criteria to reject some of the failed crops based on the generated mask size. If the size of the cropped image is less than half of the size of the original image or if the generated mask is completely blank then  we do not include it in the training  or test set. Fig~\ref{fig2} illustrates the steps of mask generating and lung ROI cropping.

For data augmentation,  2, 4, -2, and -4 degree rotations were applied and horizontal flipping was done followed by the same set of rotations. By doing so, we generated 10 times (original images, horizontal flipping, 4 sets of rotated images each from original and flipped images) more images than the original data for training. We chose a small rotation angle as X-rays are typically not rotated much.
\subsection*{Model training}
In this study, pre-trained ResNet50 \cite{he2016deep} was fine-tuned. As a base model, we used the convolutional layers pretrained on color camera images from ImageNet and removed the fully connected layers of Resnet50. Global Average pooling was applied after the last convolutional layer of the base model and a new dense layer of 64 units with ReLU activation function was added. Then, a dense layer with 1 output with sigmoid activation was added using dropout with a 0.5 probability.
All the layers of the base model were frozen during the fine-tuning procedure except the Batch Normalization layer to update the mean and variance statistics of the new dataset (X-rays). 
The total number of trainable parameters was 184K, which is approximately two orders of magnitude less than if the whole model was trained. This approach is standard procedure when training with a small dataset (transfer learning). The architecture is summarized in Table~\ref{table1}.

\begin{table}[!ht]
\centering 
\caption{
{\bf  ResNet50 Fine-Tuned Architecture.}}
\begin{tabular}{|l|}
\hline {\bf Resnet50} \\ \hline
Output from base model \\ \hline
Global Average Pooling\\ \hline
Fully Connected (64), ReLU \\ \hline
Dropout=0.5\\ \hline
Batch Normalization\\ \hline
Fully Connected (1), Sigmoid \\ \hline
Trainable parameters: 184K \\ \hline
\end{tabular}
\label{table1}
\end{table}

The model was fine-tuned using the Adam \cite{kingma2014adam} optimizer for learning with binary-cross-entropy as the loss function and a learning rate of $10^{-4}$. We set the maximum number of epochs to 200 but we stopped the training process when the validation accuracy did not improve for 5 consecutive epochs. The validation accuracy reaches its highest value of 97\% at epoch 100.

\section*{Experimental Results and Discussion}
 In our previous study \cite{hall2020finding}, we analyzed 102 Covid-19 and 102 pneumonia cases using 10-fold cross validation and a ResNet50 architecture. We obtained an overall accuracy of 89.2\% with 80.39\% of the Covid-19 cases correctly identified and other pneumonia correctly identified 101/102 for specificity/true negative rate (TNR) TNR= 0.99. Consequently, there was 1 false positive of other pneumonia. Our overall AUC was 0.95.  We note that the dataset was found to be of questionable utility.
 
In our current study, we trained the CNNs on 434 Covid-19 and 430 pneumonia chest X-ray images randomly selected from all the sources that we introduced in the previous section. For validation, 40 Covid-19 and 46 pneumonia cases were utilized. We then tested on unseen left-out data of 79 Covid-19 (30 from BIMCV Covid-19+ (Spain), 10 from Covid-19-AR (USA) and 39 from V2-COV19-NII (Germany) source) cases and 303 pneumonia (51 from NIH and 252 from Chexpert source) samples. 

For comparison purposes, we used another fine-tuning methodology where we unfroze  some of the base model convolutional layers. Thus, the weights of these layers get updated during the training process. In particular, we unfroze the last two convolutional layers of Resnet50.
We also used the two fine-tuning strategies to train another model with VGG-16 as the base model, pretrained on ImageNet.
The testing results are summarized in Table~\ref{table2}.

\begin{table}[!ht]
\centering
\caption{
{\bf  Performance results of training on a mixture of all data sources and testing on held-out test data from the same sources.} Finetune1: Freeze all base model layers, Finetune2: Unfreeze the last 2 convolutional layers. }
\begin{tabular}{|c|c|c|c|c|c|c|}
\hline
CNN & Overall Accuracy & Sensitivity & Specificity & AUC \\
\hline
\bf Resnet50-Finetune1 &  \bf 98.1\% & \bf 96.2\% & \bf 98.7\% & \bf 0.997\\
\hline
Resnet50-Finetune2 &  97\% & 95\% & 97.7\% & 0.995\\
\hline
\hline
VGG-16-Finetune1 &  88.2\% & 87.3\% & 88.4\% & 0.96 \\
\hline
VGG-16-Finetune2 & 95.5\% & 93.7\% & 96\% & 0.98 \\
\hline
\end{tabular}
\label{table2}
\end{table}

The AUC result of 0.997 seemed suspiciously high, given the difficulty expert radiologists have in diagnosing COVID-19 from CXR alone, so we decided to further investigate the robustness and generalization of these models. We evaluated the models' performances on external data sources for which there were no examples in the training data. Experiments were done with training data from just one source per class and testing data from unseen sources. Resnet-50 models with the Finetune1 method were used for the rest of the experiments in this paper. 

\begin{figure}[!h]
\caption{{\bf Overview of data splits and comparison of AUC results on seen vs. unseen test data sources.}  Note the high accuracy when held out test data is from a source included in the training set (mixing of train/test data sources). The high accuracy of these models vanishes when the data sources of the training sets are kept strictly separated from the data sources of the test sets.  }
\centering
\includegraphics[width=\textwidth]{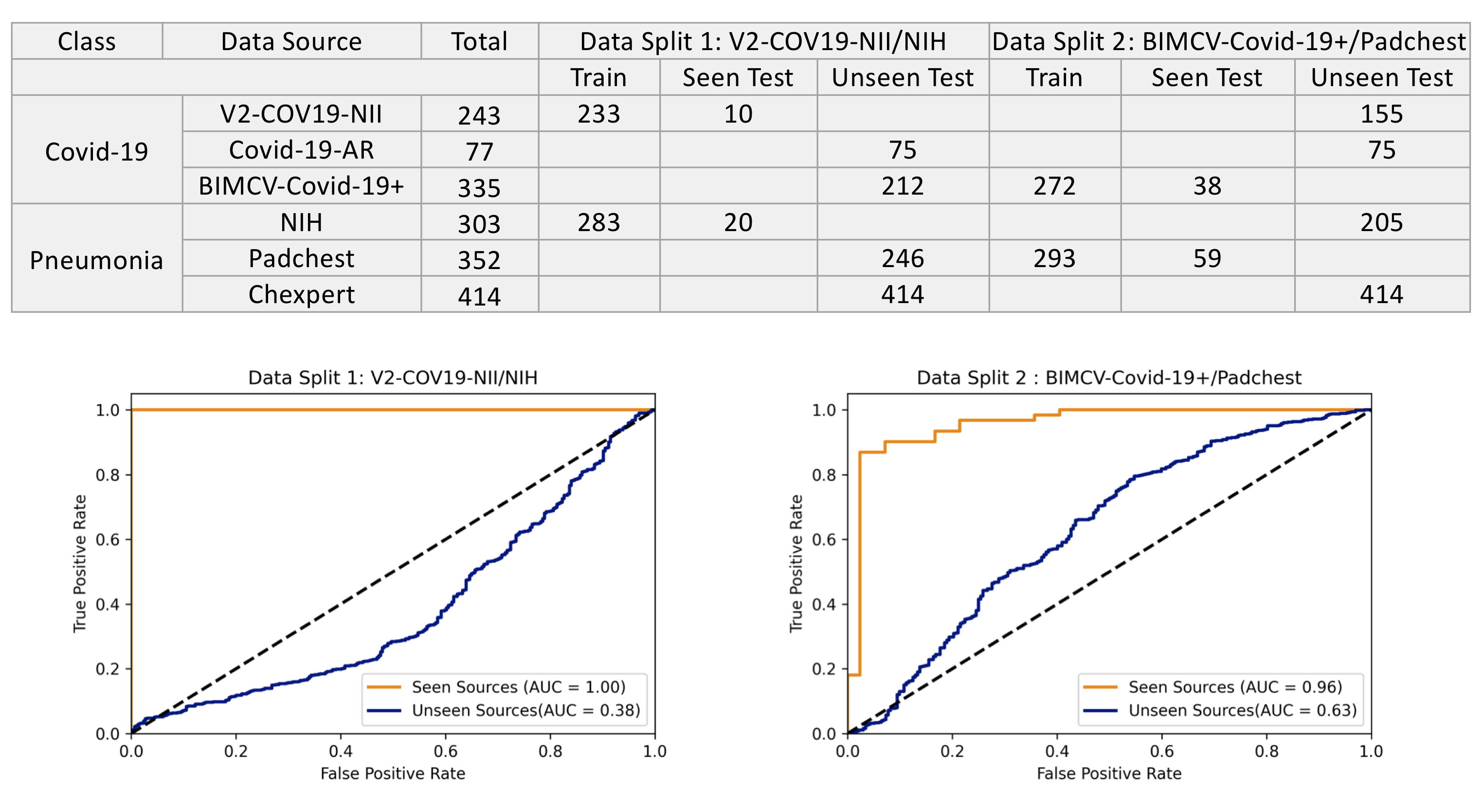}

\label{AUCexp1}
\end{figure}

As seen in the data overview table at the top of Fig. \ref{AUCexp1}, we trained the model using the V2-COV19-NII (Germany) data source for the Covid-19 class and NIH for pneumonia (Data Split 1). We then compared the AUC results on a randomly held-out subset from the seen sources (V2-COV19-NII (Germany) and NIH) versus unseen sources.

As seen in Fig. \ref{AUCexp1} to the left, the model achieves quite good results on left-out test samples from seen sources (images from the same dataset source on which the model was trained), but it performs poorly on images from unseen sources. Clearly the model was unable to generalize well to new data sources, which might indicate that the model is relying on confounding information related to the data sources instead of the real underlying pathology of Covid-19. The fact that its performance (AUC=0.38) is less than AUC=0.5 (worse than random), strongly suggests that the model is relying on confounding information. The perfect score on the data from the seen dataset source also hints at confounders, as it is unlikely that any algorithm could perfectly distinguish Covid-19 positive versus pneumonia patients based on lung findings alone. On the other hand, it is highly likely that perfect classification could be performed based on the images data-source. To give a human analogy, a radiologist would find it easier to classify Covid-19+ versus Covid-19-negative chest X-rays by looking at the year in which the image was taken (pre-2020 versus post), rather than by looking at the image itself.

As presented in the data overview table on top of Fig. \ref{AUCexp1}, in an experiment to see if a model built with data from similar sources resulted in more general models, we chose a second data split (data split 2) with BIMCV-Covid-19+ (Spain) data as the source for Covid-19 and Padchest for Pneumonia. These two sources come from the same regional healthcare system (Valencia, Spain), both were prepared by the same team and underwent the same data pre-processing. We anticipated that reducing the differences between classes in terms of image normalization, hospitals, scanners, image acquisition protocols, etc would enable the model to only concentrate on learning medically-relevant markers of Covid-19 instead of source specific confounders.

The results in Fig \ref{AUCexp1} to the right show that the model still exhibits high performance on seen sources but generalizes poorly to external sources. Therefore, we can see that even having both classes from the same hospital system did not prevent the model from learning data-source specific confounders. However, in contrast to the model trained on Data Split1, this model has worse performance on data from seen sources (AUC=0.96 vs AUC=1.00) and better performance on data from unseen sources (AUC=0.63 vs AUC=0.38). Notably, the second model's performance is better than random (AUC$>$0.5). This suggests that the algorithm may have actually learned some clinically salient features, although once again, the majority of its performance appears to be  based on confounders.

We can also observe that it is possible that confounders found in some data sources can generalize across sources. For example when training using the BIMCV-Covid-19+ (Spain) data source, the model had an accuracy of 88\% on Covid-19-AR (USA), which is an unseen source. However when training using V2-COV19-NII (Germany) data source, the model only achieved an accuracy of 68\% on this same unseen source (Covid-19-AR (USA)).

As a possible solution, we tried fine-tuning the trained model from the experiment using multiple sources for each class, using a subset of 80 samples from BIMCV-Covid-19+ (Spain) for the Covid-19 class and a subset of 80 samples from Chexpert for the pneumonia class. Both these  sources were considered unseen in the experiment with data split1 described in the data overview table on top of Fig. \ref{AUCexp1}.
As seen in Table \ref{finetune}, fine-tuning with subsets from unseen sources improves the model's overall performance on those sources. We hypothesize that fine-tuning helps the model to ignore noisy features and data-source related confounders and instead concentrate on learning meaningful and robust features.
To investigate what the model is actually relying on this time, we applied the GradCam algorithm \cite{selvaraju2017grad} to test samples and highlight areas in the image that have a strong effect on classification. We would expect a classifier relying on true pathologic features to primarily be relying on pixels from the lung fields, whereas a spurious classifier would rely on pixels from regions of the image irrelevant to diagnosis. The results were inconclusive. Therefore, we cannot affirm whether the model is still relying on shortcuts/confounders to make decisions. This experimental result shows that a model could be adapted to work locally. Still to be shown is that it learns medically relevant features. 

\begin{table}[!ht]
\centering
\caption{
{\bf Accuracy results of finetuning a model built on multiple sources from both classes to adapt it to work locally.} Still to be shown is that it learns medically relevant features.  }
\begin{tabular}{|c|c|c|c|}
\hline 
Class & Data Source & Before & After  \\ 
 \hline
 \multirow{1}{*}{Covid-19} & BIMCV-Covid-19+ (Spain) & 51\% & 98\%  \\

\hline

\multirow{1}{*}{Pneumonia} & Chexpert & 12\% & 94\% \\
 \hline
\end{tabular}
\label{finetune}
\end{table}

\section*{Conclusions}

In this paper we demonstrate that deep learning models can leverage data-source specific confounders to differentiate between Covid-19 and pneumonia labels. While we eliminated many confounders from earlier work, such as those related to large age discrepancies between populations (pediatric vs adult), image post-processing artifacts introduced by working from low resolution pdf images, and positioning artifacts by pre-segmenting and cropping the lungs, we still saw that deep-learning models were able to learn using data-source specific confounders. Several hypotheses may be considered as to the nature of these confounders. These confounders may be introduced as a result of differences in X-ray procedures as a result of patient clinical severity or patient control procedures. For instance, differences in disease severity may impact patient positioning (standing for ambulatory or emergency department patients vs supine for admitted and ICU patients). In addition, if a particular X-ray machine whose signature is learnable is always used for Covid-19 patients, because it is in a dedicated Covid-19 ward, this would be another method to determine the class in a non-generalizable way.

Using datasets that underwent different pre-processing methods across classes can encourage the model to differentiate classes based on the pre-processing, which is an undesirable outcome. Thus, training the model on a dataset of raw data coming from many sources may provide a general classifier. Even within the same hospital, one must still check to be sure that something approximating what a human would use to differentiate cases is learned. 

That being said, using a deep learning classifier trained on positive and negative datasets from the same hospital system, having undergone similar data processing, we were able to train a classifier that performed better than random on chest X-rays from unseen data sources, albeit modestly. This suggests that this classification problem may eventually be solvable using deep learning models. However, the theoretical limit of Covid-19 diagnosis, based solely on chest X-ray remains unknown, and consequently also the maximum expected AUC of any machine learning algorithm. Unlike other classification problems that we know can be performed with high accuracy by radiologists, radiologists do not routinely or accurately diagnose Covid-19 by chest X-ray alone. However, an imperfect classifier that has learned features that are not confounders can be combined with other clinical data to create a highly accurate classifiers, and as such this area warrants further inquiry. 

Our results suggest that, for at least this medical imaging problem, when deep learning is involved it is important to have data from unseen sources (pre-processed in the same way) included in a test set.  If there are no unseen sources available, careful investigation is necessary to ensure that what is learned is both generalizable and germane.

\section*{Data availability}

All data and code used in this study are available in our Github repository at: 
\href{https://github.com/kbenahmed89/Pretrained-CNN-For-Covid-19-Prediction-from-Automatically-Lung-ROI-Cropped-X-Rays}{github.com-USF-COVID}.

%
%
%


\begin{thebibliography}{10}

\bibitem{worldometers}
worldometers. worldometers COVID-19 CORONAVIRUS PANDEMIC; 2020 (accessed Dec
  22, 2020).
\newblock Available from: \url{https://www.worldometers.info/coronavirus/}.

\bibitem{weissleder2020Covid}
Weissleder R, Lee H, Ko J, Pittet MJ.
\newblock COVID-19 diagnostics in context.
\newblock Science Translational Medicine. 2020;12(546).

\bibitem{jama}
Wang W, Xu Y, Gao R, Lu R, Han K, Wu G, et~al.
\newblock {Detection of SARS-CoV-2 in Different Types of Clinical Specimens}.
\newblock JAMA. 2020;doi:{10.1001/jama.2020.3786}.

\bibitem{stopsign}
Eykholt K, Evtimov I, Fernandes E, Li B, Rahmati A, Xiao C, et~al.
\newblock Robust physical-world attacks on deep learning visual classification.
\newblock In: Proceedings of the IEEE Conference on Computer Vision and Pattern
  Recognition; 2018. p. 1625--1634.

\bibitem{chen2020survey}
Chen Y, Jiang G, Li Y, Tang Y, Xu Y, Ding S, et~al.
\newblock A Survey on Artificial Intelligence in Chest Imaging of COVID-19.
\newblock BIO Integration. 2020;.

\bibitem{ozturk2020classification}
Ozturk S, Ozkaya U, Barstugan M.
\newblock Classification of coronavirus images using shrunken features.
\newblock medRxiv. 2020;.

\bibitem{abbas2020classification}
Abbas A, Abdelsamea MM, Gaber MM.
\newblock Classification of COVID-19 in chest X-ray images using DeTraC deep
  convolutional neural network.
\newblock arXiv preprint arXiv:200313815. 2020;.

\bibitem{farooq2020Covid}
Farooq M, Hafeez A.
\newblock Covid-resnet: A deep learning framework for screening of covid19 from
  radiographs.
\newblock arXiv preprint arXiv:200314395. 2020;.

\bibitem{lv2020cascade}
Lv D, Qi W, Li Y, Sun L, Wang Y.
\newblock A cascade network for Detecting COVID-19 using chest x-rays.
\newblock arXiv preprint arXiv:200501468. 2020;.

\bibitem{bassi2020deep}
Bassi PR, Attux R.
\newblock A Deep Convolutional Neural Network for COVID-19 Detection Using
  Chest X-Rays.
\newblock arXiv preprint arXiv:200501578. 2020;.

\bibitem{rahimzadeh2020new}
Rahimzadeh M, Attar A.
\newblock A New Modified Deep Convolutional Neural Network for Detecting
  COVID-19 from X-ray Images.
\newblock arXiv preprint arXiv:200408052. 2020;.

\bibitem{chowdhury2020can}
Chowdhury ME, Rahman T, Khandakar A, Mazhar R, Kadir MA, Mahbub ZB, et~al.
\newblock Can AI help in screening viral and COVID-19 pneumonia?
\newblock arXiv preprint arXiv:200313145. 2020;.

\bibitem{hemdan2020Covidx}
Hemdan EED, Shouman MA, Karar ME.
\newblock Covidx-net: A framework of deep learning classifiers to diagnose
  covid-19 in x-ray images.
\newblock arXiv preprint arXiv:200311055. 2020;.

\bibitem{karim2020deepCovidexplainer}
Karim M, D{\"o}hmen T, Rebholz-Schuhmann D, Decker S, Cochez M, Beyan O, et~al.
\newblock Deepcovidexplainer: Explainable covid-19 predictions based on chest
  x-ray images.
\newblock arXiv preprint arXiv:200404582. 2020;.

\bibitem{hall2020finding}
Hall LO, Paul R, Goldgof DB, Goldgof GM.
\newblock Finding covid-19 from chest x-rays using deep learning on a small
  dataset.
\newblock arXiv preprint arXiv:200402060. 2020;.

\bibitem{Apostolopoulos_2020}
Apostolopoulos ID, Mpesiana TA.
\newblock Covid-19: automatic detection from X-ray images utilizing transfer
  learning with convolutional neural networks.
\newblock Physical and Engineering Sciences in Medicine.
  2020;doi:{10.1007/s13246-020-00865-4}.

\bibitem{apostolopoulos2020extracting}
Apostolopoulos ID, Aznaouridis SI, Tzani MA.
\newblock Extracting possibly representative COVID-19 Biomarkers from X-Ray
  images with Deep Learning approach and image data related to Pulmonary
  Diseases.
\newblock Journal of Medical and Biological Engineering. 2020; p.~1.

\bibitem{basu2020deep}
Basu S, Mitra S.
\newblock Deep Learning for Screening COVID-19 using Chest X-Ray Images.
\newblock arXiv preprint arXiv:200410507. 2020;.

\bibitem{li2020robust}
Li T, Han Z, Wei B, Zheng Y, Hong Y, Cong J.
\newblock Robust Screening of COVID-19 from Chest X-ray via Discriminative
  Cost-Sensitive Learning.
\newblock arXiv preprint arXiv:200412592. 2020;.

\bibitem{yeh2020cascaded}
Yeh CF, Cheng HT, Wei A, Liu KC, Ko MC, Kuo PC, et~al.
\newblock A Cascaded Learning Strategy for Robust COVID-19 Pneumonia Chest
  X-Ray Screening.
\newblock arXiv preprint arXiv:200412786. 2020;.

\bibitem{panwar2020application}
Panwar H, Gupta P, Siddiqui MK, Morales-Menendez R, Singh V.
\newblock Application of Deep Learning for Fast Detection of COVID-19 in X-Rays
  using nCOVnet.
\newblock Chaos, Solitons \& Fractals. 2020; p. 109944.

\bibitem{kassani2020automatic}
Kassani SH, Kassasni PH, Wesolowski MJ, Schneider KA, Deters R.
\newblock Automatic Detection of Coronavirus Disease (COVID-19) in X-ray and CT
  Images: A Machine Learning-Based Approach.
\newblock arXiv preprint arXiv:200410641. 2020;.

\bibitem{oh2020deep}
Oh Y, Park S, Ye JC.
\newblock Deep learning covid-19 features on cxr using limited training data
  sets.
\newblock IEEE Transactions on Medical Imaging. 2020;.

\bibitem{wang2020Covid}
Wang L, Wong A.
\newblock COVID-Net: A Tailored Deep Convolutional Neural Network Design for
  Detection of COVID-19 Cases from Chest X-Ray Images.
\newblock arXiv preprint arXiv:200309871. 2020;.

\bibitem{Cherezov-Size}
Cherezov D, Paul R, Fetisov N, Gillies RJ, Schabath MB, Goldgof DB, et~al.
\newblock Lung Nodule Sizes Are Encoded When Scaling CT Image for CNN's.
\newblock Tomography. 2020;6(2):209--215.

\bibitem{degrave2020ai}
DeGrave AJ, Janizek JD, Lee SI.
\newblock AI for radiographic COVID-19 detection selects shortcuts over signal.
\newblock medRxiv. 2020;.

\bibitem{cohen2020Covid}
Cohen JP, Morrison P, Dao L.
\newblock COVID-19 image data collection.
\newblock arXiv preprint arXiv:200311597. 2020;.

\bibitem{radiopedia}
Radiopedia Covid data :https://radiopaedia.org/. 2020;.

\bibitem{sirm}
SIRM Covid data :http://www.sirm.org/en/. 2020;.

\bibitem{de2020bimcv}
de~la Iglesia~Vay{\'a} M, Saborit JM, Montell JA, Pertusa A, Bustos A, Cazorla
  M, et~al.
\newblock BIMCV COVID-19+: a large annotated dataset of RX and CT images from
  COVID-19 patients.
\newblock arXiv preprint arXiv:200601174. 2020;.

\bibitem{tcia_2020}
Desai S, Baghal A, Wongsurawat T, Al-Shukri S, Gates K, Farmer P, et~al.. Chest
  Imaging with Clinical and Genomic Correlates Representing a Rural COVID-19
  Positive Population [Data set]; 2020.
\newblock Available from:
  \url{https://wiki.cancerimagingarchive.net/pages/viewpage.action?pageId=70226443#702264434dc5f53338634b35a3500cbed18472e0}.

\bibitem{germanymeyer_2020}
Winther HB, Laser H, Gerbel S, Maschke SK, B~Hinrichs J, Vogel-Claussen J,
  et~al.. COVID-19 Image Repository; 2020.
\newblock Available from:
  \url{https://figshare.com/articles/dataset/COVID-19_Image_Repository/12275009/1}.

\bibitem{wang2017chestx}
Wang X, Peng Y, Lu L, Lu Z, Bagheri M, Summers RM.
\newblock Chestx-ray8: Hospital-scale chest x-ray database and benchmarks on
  weakly-supervised classification and localization of common thorax diseases.
\newblock In: Proceedings of the IEEE conference on computer vision and pattern
  recognition; 2017. p. 2097--2106.

\bibitem{irvin2019chexpert}
Irvin J, Rajpurkar P, Ko M, Yu Y, Ciurea-Ilcus S, Chute C, et~al.
\newblock Chexpert: A large chest radiograph dataset with uncertainty labels
  and expert comparison.
\newblock In: Proceedings of the AAAI Conference on Artificial Intelligence.
  vol.~33; 2019. p. 590--597.

\bibitem{bustos2020padchest}
Bustos A, Pertusa A, Salinas JM, de~la Iglesia-Vay{\'a} M.
\newblock Padchest: A large chest x-ray image dataset with multi-label
  annotated reports.
\newblock Medical image analysis. 2020;66:101797.

\bibitem{bustos2019padchest}
Bustos A, Pertusa A, Salinas JM, de~la Iglesia-Vay{\'a} M.
\newblock Padchest: A large chest x-ray image dataset with multi-label
  annotated reports.
\newblock arXiv preprint arXiv:190107441. 2019;.

\bibitem{sivaramakrishnan2004iteratively}
Sivaramakrishnan R, Jen S, Philip O, et~al.
\newblock Iteratively Pruned Deep Learning Ensembles for COVID-19 Detection in
  Chest X-rays.
\newblock arXiv preprint arXiv:200408379;.

\bibitem{he2016deep}
He K, Zhang X, Ren S, Sun J.
\newblock Deep residual learning for image recognition.
\newblock In: Proceedings of the IEEE conference on computer vision and pattern
  recognition; 2016. p. 770--778.

\bibitem{kingma2014adam}
Kingma DP, Ba J.
\newblock Adam: A method for stochastic optimization.
\newblock arXiv preprint arXiv:14126980. 2014;.

\bibitem{selvaraju2017grad}
Selvaraju RR, Cogswell M, Das A, Vedantam R, Parikh D, Batra D.
\newblock Grad-cam: Visual explanations from deep networks via gradient-based
  localization.
\newblock In: Proceedings of the IEEE international conference on computer
  vision; 2017. p. 618--626.

\end{thebibliography}

\section*{Acknowledgments}
This research was partially supported by NSF award (1513126). Gregory Goldgof was partially supported by the NIH (R38HL143581).
We would like to thank Sudheer Nadella for his contribution in the lung cropping task.

\section*{Author contributions}
K.B.A. performed the experimental work and prepared the manuscript. L.O.H. and D.G. provided project leadership and manuscript review. G.G. developed the theory and assisted with manuscript writing. R.P. participated in the manuscript writing.

\section*{Competing interests}
No conflicts of interest, financial or otherwise, are declared by the authors.

\end{document}